\documentclass{article}

\usepackage{arxiv}
\usepackage{siunitx} % For proper unit formatting
\usepackage[utf8]{inputenc} % allow utf-8 input
\usepackage[T1]{fontenc}    % use 8-bit T1 fonts
\usepackage{hyperref}       % hyperlinks
\usepackage{url}            % simple URL typesetting
\usepackage{booktabs}       % professional-quality tables
\usepackage{amsfonts}       % blackboard math symbols
\usepackage{nicefrac}       % compact symbols for 1/2, etc.
\usepackage{microtype}      % microtypography
\usepackage{lipsum}
\usepackage{caption}
\usepackage{amsmath}
\usepackage{graphicx}
\usepackage{natbib}
\graphicspath{ {./images/} }

\title{Distinguishing Initial-Condition and Model Error Using Machine-Learning Weather Models}

\usepackage{authblk}

\newbox{\orcid}
\sbox{\orcid}{\includegraphics[scale=0.06]{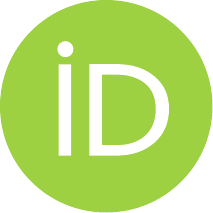}}

\author[1,2]{%
    \href{https://orcid.org/0009-0006-7242-2351}{%
    \usebox{\orcid}\hspace{1mm}P. Trent Vonich}%
}

\author[1]{%
    \href{https://orcid.org/0000-0001-8486-9739}{%
    \usebox{\orcid}\hspace{1mm}Gregory J. Hakim}%
}

\affil[1]{Department of Atmospheric Sciences, University of Washington, Seattle, WA, USA}
\affil[2]{Air Force Institute of Technology, Wright-Patterson AFB, OH, USA}

\affil[ ]{\small
    \texttt{tvonich@uw.edu}
    \qquad
    \texttt{ghakim@uw.edu}
}

\begin{document}
\maketitle

 \begin{abstract}
Differentiable weather models allow atmospheric initial conditions to be directly optimized to shadow a known trajectory, which we take from a reanalysis dataset. Here, we extend this optimization to include the forecast model, GraphCast, using low-rank adaptation (LoRA). We compare model optimization with initial-condition (IC) optimization, and then joint optimization, where both are tuned simultaneously. When compared against a control forecast initialized with reanalysis data, LoRA and IC optimization individually produce similar improvements. We find similar results when IC and LoRA are jointly optimized across a spectrum of solutions where a tuning coefficient smoothly shifts the weight of the joint optimization from all-LoRA to all-IC; forecast improvement is not unique to the relative contributions of model and IC corrections. To better discriminate the source of error, we introduce a second model, Pangu-Weather, as a supervisor that gates IC updates during joint optimization. This gated optimization is the same as before, except that IC updates at each epoch are accepted only when they improve the forecasts of both models; otherwise, only the LoRA update is accepted. Results show that this cross-model constraint substantially improves agreement between the two models in their response to the same optimized ICs, narrowing the gap in mean forecast error reduction from 64\% to 10\%. Moreover, when used together, the LoRA and IC corrections from the gated optimization yield GraphCast forecast improvements similar to those from the ungated optimization. These results imply that model error that is encoded in the ungated ICs is shifted to the LoRA corrections via gating, yielding IC corrections that may provide better analysis estimates. 
\end{abstract}

% % % % % % % % % % % % % % % % % % % % % % % % % % % % % % % % % % % % % % % % % % % % % % % % %
  
\section{Introduction}\label{sec:intro}
Forecast error in numerical weather prediction reflects both imperfect knowledge of the atmospheric state and deficiencies in the model that advances it. Optimizing the initial condition against later observations, as in strong-constraint four-dimensional variational data assimilation (4DVAR), can reduce forecast error by encoding observational information into the initial state \citep[e.g.,][]{TalagrandCourtier1987, Pires1996}. The resulting reduction in forecast error, however, does not reveal how much of the improvement derives from a more accurate initial state versus compensation for model error. The computational efficiency and end-to-end differentiability of machine-learning (ML) models enable renewed investigation of this well-established attribution problem by making it possible to search initial states over previously impractical forecast ranges and to include model corrections as well.

Initial-condition (IC) optimization in ML models illustrates both the promise and the challenge of associating forecast improvements with the IC modifications. Using GraphCast \citep{lam2023learning}, \citet{vonich2026} employ IC optimization to obtain an average 86\% reduction in ten-day forecast error across 732 initialization times, with forecast skill extending to $\sim$30 days. When the GraphCast-optimized initial conditions are used to initiate Pangu-Weather \citep{bi2023}, however, the average error reduction peaks at only 21\% on day 4. This contrast motivates the present study, as the optimized increment may correct analysis error while also compensating for model error specific to GraphCast. This attribution problem is well established in strong-constraint 4DVAR, where improved forecast fit need not---and often does not---imply a more accurate estimate of the initial state \citep{isaksen_2005,tothpena2007}. 

The ambiguity in attributing forecast error to the initial condition versus the model has a long history. Some studies separate initial-condition and model-error contributions by assuming that state error grows from initialization, while model deficiencies continuously add error during the forecast \citep{dalcher1987error, tothpena2007}. \citet{Swanson-1998} show that model error can limit the benefits of long assimilation windows, eventually degrading both state estimates and forecasts. In a complementary line of work, the adjoint-based ``key analysis error'' framework, analogous to the optimization approach of \citet{heatwave_2024}, seeks the initial perturbations that  most reduces 48-hour forecast error using tangent linear and adjoint models \citep{klinker_1996,Klinker1998}. Although successful at reducing forecast error, \citet{isaksen_2005} established that such perturbations cannot be uniquely interpreted as analysis corrections. Different verification norms produce markedly different perturbation structures despite comparable forecast improvements, and the resulting perturbations often degraded the first 12 hours of the forecast. \citet{orrell2001model} identify a more fundamental limit to the key analysis framework, showing that model error sets an approximate lower bound on how closely any model trajectory can shadow the truth, such that small perturbations to the initial condition cannot offset model error beyond a limited time.

Weak-constraint 4DVAR aims to make the attribution problem more explicit by including both the initial state and model error as control variables \citep[e.g.,][]{fisher2005equivalence,fisher2011long,fisher2011weak,tremolet2007model}. \citet{juddsmith2004} show that even small model imperfections can substantially bias state estimates when the model is assumed to be perfect. They propose using pseudo-orbits, allowing departures from the model's predicted trajectory to account for errors in its dynamics. This adaptability does not, however, fully resolve the underlying attribution ambiguity, as corrections to the initial condition and model can compensate for one another in minimizing the forecast objective \citep{tremolet2006accounting,tremolet2007modelerror}. Operational data assimilation systems delineate the two main error types through structural assumptions about their properties such as spatial scale, temporal persistence, and prior covariance \citep{laloyaux2020exploring,bonavita2021timecorrelated}. Such \textit{a priori} divisions are useful only to the extent that the assumed structures match the actual errors. More recent work has explored learned model-error corrections, progressing from offline hybrid-model approaches to formulations embedded directly within weak-constraint 4DVAR \citep{farchi2021using,farchi2023online,farchi2025development}.

Yet even with more flexible representations of model error, attribution remains challenging when both state and model corrections are inferred within a single forecast system. Transplant experiments, in which one forecast model is initialized from another system's analysis, provide a more direct test of whether forecast differences stem from the analysis or the model. Initially limited to case studies, this approach was later extended through broader intermodel and observing-system experiments \citep{harrison1999a,magnusson2019dependence}. These studies generally associate less systematic, case-dependent forecast errors with initial-condition differences and persistent biases with model formulation differences. At a more basic level, \citet{prive2013role} show that data assimilation embeds systematic model errors in the analysis where observations are sparse, carrying them into the initial conditions for the subsequent forecast.

In this study, we show that differentiable ML weather models do not resolve this classical degeneracy, but offer efficient new ways to interrogate it through direct access to the neural-network \citep{toms2020physically,explainable_2024}. Direct access to parameters lends itself to a targeted model correction. As the counterpart to IC optimization \citep{vonich2026}, we use low-rank adaptation (LoRA) to optimize the forecast model itself for a given trajectory. Originally developed as a parameter-efficient method for post-training large language models \citep{LoRA_in_LLMs}, LoRA learns low-rank changes to the base model, with rank providing direct control over the dimensionality of the correction. The Aurora model employs this technique, making it tractable to train against a 10-day forecast loss \citep{Aurora}. Note that weight correction through LoRA differs from the additive model-error forcing used in traditional weak-constraint 4DVAR because it modifies the forecast operator and remains fixed throughout the rollout, rather than applying a separate correction to the evolving model state \citep[e.g.,][]{tremolet2006accounting,fisher2011weak}. We use it here as a parsimonious control on model error that can be optimized against the same verification target as an initial-condition increment.

We explore the separation of initial-condition and model error in four stages using JAX automatic differentiation \citep{jax2018github}. First, we hold the initial condition fixed and optimize only a LoRA weight correction, $\delta W$, through GraphCast, to compare how closely a low-rank model correction can shadow the verifying trajectory relative to initial condition optimization. Second, we simultaneously optimize an IC correction, $\delta x$, and a model correction, $\delta W$. A scalar regularization term varies the relative penalty on the two controls, allowing us to examine how that balance, together with optimization window-length, affects performance in a different model, Pangu-Weather. Next, we use Pangu-Weather as a supervisor to gate the optimization in a proof-of-concept experiment aimed at separating IC and model error. After each simultaneous IC--LoRA gradient calculation, the IC update is retained only if forecasts improve for both Pangu-Weather and GraphCast; if Pangu-Weather does not improve, only the LoRA update may be committed. This tests whether an independent forecast model can guide the state optimization toward IC updates that are beneficial for both models. Last, we extend model optimization by training a separate LoRA correction for each 24-hour period of a 10-day forecast without consideration of the IC. This introduces time-dependent model adjustments, conceptually similar to the tendency corrections used in traditional weak-constraint 4DVAR.

Together, these experiments compare the state-only and model-only limits of forecast optimization and probe the range of state--model error partitions that arise when both controls are available. Section~\ref{sec2:data_methods} describes the methodology for the four experiment types. Section~\ref{sec3:results} covers results, including forecast dependence on LoRA rank, the full range of state--model weighting experiments, and using Pangu-Weather as a supervisor on the initial condition corrections. The time-dependent LoRA results are described in Section~\ref{results:multilora}. Section~\ref{sec5:conclusion} provides a concluding discussion and ideas for future work.

\section{Data and Methods}
\label{sec2:data_methods}
All experiments use the 1.0$^\circ$ version of GraphCast \citep{lam2023learning} implemented in JAX, with ECMWF Reanalysis v5 (ERA5) providing the initial atmospheric states and serving as the verification reference. The 1.0$^\circ$ configuration is used for its low computational cost and competitive forecast performance at the spatial scales considered here. The experiments begin by comparing IC-only and LoRA-only optimization, followed by joint optimization with variable weighting to examine sensitivity to the relative contributions of IC and model corrections. Pangu-Weather (0.25$^\circ$) is used to provide an independent estimate of forecast skill from IC updates, allowing it to gate their acceptance during optimization. Finally, we study the effect of independent LoRA model corrections at successive forecast time steps. The baseline IC optimization framework from \citet{vonich2026} is summarized in Appendix~\ref{appendix:baseline_methods}, while the new methods are described below. Throughout the remainder of the paper, ``forecast improvement'' denotes a reduction in forecast error between a test case and a control, both verified against ERA5, where the control forecast uses ERA5 initial conditions.

\subsection{LoRA Model-Weight Optimization}
\label{lora_only}
 \citet{heatwave_2024} and \citet{vonich2026} investigate how long GraphCast can shadow ERA5 when only the initial conditions are adjusted. Here, we examine the complementary question: how long can ERA5 be shadowed when only the forecast model is allowed to change? LoRA offers a natural framework for this experiment because it enables targeted fine-tuning of the model while leaving the original GraphCast weights fixed and provides direct control over the dimensionality of the model corrections.

GraphCast comprises an encoder, a processor, and a decoder. We target the 16 neural-network layers of the processor, each of which contains an edge-update multilayer perceptron (MLP) and a node-update MLP. Each MLP contains two learned weight matrices, giving four matrices per layer and 64 targeted matrices across the processor. Unless otherwise specified, we apply LoRA to all 64 matrices. For the \textit{k}-th targeted matrix, we define

$$
\mathbf{W}_k \in \mathbb{R}^{m_k \times n_k},
$$

where $k$ indexes the targeted matrix and $m_k$ and $n_k$ are its numbers of rows and columns, respectively. Rather than directly optimizing the $m_k n_k$ elements of $\mathbf{W}_k$, LoRA represents its correction using two smaller trainable matrices,

$$
\mathbf{B}_k \in \mathbb{R}^{m_k \times r},
\qquad
\mathbf{A}_k \in \mathbb{R}^{r \times n_k},
$$

where $r$ is the LoRA rank. Their product has dimensions

$$
(m_k \times r)(r \times n_k)
=
m_k \times n_k,
$$

matching the shape of the original weight matrix $\mathbf{W}_k$. This factorization uses $r(m_k+n_k)$ trainable parameters instead of $m_kn_k$. As long as $r \ll \min(m_k,n_k)$, the LoRA correction occupies a much lower-dimensional parameter space than the original weight matrix.

For each targeted matrix, we define the LoRA correction as

\begin{equation}
\delta\mathbf{W}_k^{(r)}
=
\frac{\alpha}{\sqrt{r}}\mathbf{B}_k\mathbf{A}_k ,
\label{eq:lora_update}
\end{equation}

Before each forward pass, this correction is added to the original GraphCast weight matrix,

\begin{equation}
\mathbf{W}_k^{\mathrm{eff}}
=
\mathbf{W}_k+\delta\mathbf{W}_k^{(r)},
\label{eq:lora_effective_weight}
\end{equation}

and $\mathbf{W}_k^{\mathrm{eff}}$ is used in place of $\mathbf{W}_k$. The original GraphCast weights remain fixed throughout optimization; only $\mathbf{A}_k$ and $\mathbf{B}_k$ are updated.

We use the rank-stabilized LoRA scaling factor $\alpha/\sqrt{r}$ to reduce rank-dependent changes in adapter magnitude and gradient scale \citep{kalajdzievski2023rank,wang2025block}. Following the original LoRA convention of setting $\alpha$ to a reference rank and holding it fixed as rank varies \citep{LoRA_in_LLMs}, we set $\alpha=2$. 

The $\mathbf{B}_k$ matrices are initialized to zero, while the elements of $\mathbf{A}_k$ are drawn independently from a zero-mean Gaussian distribution with standard deviation $0.01/\sqrt{r}$. The constant $0.01$ sets a small initialization scale and was not tuned. This gives $\delta\mathbf{W}_k^{(r)}=0$ at initialization while still allowing gradients to update $\mathbf{B}_k$ immediately; if both factors were initialized to zero, their gradients would vanish. Each optimization therefore begins from the unmodified GraphCast model.

For a given experiment, the total number of trainable LoRA parameters is

\begin{equation}
N_{\mathrm{par}}(r)
=
r\sum_{k\in\mathcal{K}}(m_k+n_k),
\label{eq:lora_npar}
\end{equation}

where $\mathcal{K}$ denotes the set of targeted weight matrices. 

The LoRA parameters are optimized against the same weighted forecast MSE used for initial-condition optimization, while the ERA5 initial condition is held fixed. Optimization uses the same progressive 2-day lead-time expansion procedure used in \citet{vonich2026}, carrying the optimized LoRA parameters forward as the forecast window is extended (50 epochs per window). Both the loss function and the optimization procedure are described in Appendix~\ref{appendix:baseline_methods}. Gradients are computed through the autoregressive GraphCast forecast using JAX, and $\mathbf{A}_k$ and $\mathbf{B}_k$ are updated using the Adam optimizer.

\subsection{Joint Optimization of the Initial Condition and Model Weights}
\label{sec:joint}

The joint experiment allows both the initial-condition increment $\delta\mathbf{x}$ and the LoRA weight correction $\delta\mathbf{W}$ to be optimized against the same forecast objective. Rank is selected at $r=2$ based on the LoRA-only results described in section \ref{sec:lora_ladder}. Both corrections are zero at initialization, so each optimization begins from the unmodified ERA5 analysis and GraphCast model.

We minimize

\begin{equation}
\mathcal{J}(\delta\mathbf{x},\delta\mathbf{W})
=
\mathcal{L}_w
+
\lambda_{\mathrm{IC}}\widetilde{S}_{\mathrm{IC}}
+
\lambda_W\widetilde{S}_W
+
\mu R_{\nabla^2}.
\label{eq:joint_objective}
\end{equation}

Here, $\mathcal{L}_w$ is the GraphCast forecast loss over the current optimization window, as defined in Appendix~\ref{appendix:baseline_methods}. The next two terms penalize the magnitudes of the initial-condition and model-weight corrections, respectively, allowing us to vary their relative contribution. Because the IC and LoRA controls differ in shape and natural magnitude, we use dimensionless normalized penalties,

\begin{equation}
\widetilde{S}_{\mathrm{IC}}
=
\frac{S_{\mathrm{IC}}}{S_{\mathrm{IC}}^{\mathrm{ref}}},
\qquad
\widetilde{S}_W
=
\frac{S_W}{S_W^{\mathrm{ref}}}.
\label{eq:joint_normalized_norms}
\end{equation}

Here, $S_{\mathrm{IC}}^{\mathrm{ref}}$ and $S_W^{\mathrm{ref}}$ are taken from that date's joint solution obtained without penalty terms and held fixed across the subsequent penalty sweep. This places the two penalties on a comparable dimensionless reference scale to account for the different control space sizes.

The corresponding raw magnitudes are

\begin{equation}
S_{\mathrm{IC}}
=
\left\lVert \delta\mathbf{x}_{\mathrm{std}} \right\rVert_{\cos\phi}^{2},
\qquad
S_W
=
\sum_k
\left\lVert\delta\mathbf{W}_k\right\rVert_F^2.
\label{eq:joint_norms}
\end{equation}

$\delta\mathbf{x}_{\mathrm{std}}$ denotes the standardized full-state IC correction across all optimized variables, vertical levels, input times, and grid points. The $\cos\phi$ weighting accounts for the decrease in grid-cell area toward the poles. For $S_W$, $k$ indexes the LoRA-targeted weight matrices, and $\lVert\cdot\rVert_F$ denotes the Frobenius norm.

The final term in Eq.~\ref{eq:joint_objective} is an augmentation of the method of \citet{vonich2026} that penalizes small-scale noise in the initial-condition correction. Small-scale noise can appear in optimized initial conditions because ML weather models are potent denoisers \citep{denoiser}. Unlike the LoRA correction, which acts at every autoregressive step, the IC is constrained by the forecast loss only through its effect on subsequent states. Small-scale IC perturbations that are rapidly damped during the first step may therefore remain largely invisible to the loss. We impose the Laplacian penalty directly on the physical-space grid. $R_{\nabla^2}$ is the $\cos\phi$-weighted sum of the squared five-point discrete Laplacian of $\delta\mathbf{x}$, with equal weighting across variables. We set $\mu=5\times10^{-4}$ based on empirical experimentation.

A single backward pass through GraphCast produces gradients with respect to both controls using JAX automatic differentiation \citep{jax2018github}. Distinct Adam optimizers, both using standard hyperparameters and an initial learning rate of $5\times10^{-4}$, update $\delta\mathbf{x}$ and the LoRA parameters. Optimization proceeds through forecast windows of 2, 4, 6, 8, and 10 days, with 30 iterations at each window before proceeding to the next. At each window, both optimizers are re-initialized with a learning rate scaled by $1/T_w$, where $T_w$ is the window length. This prevents the loss from overshooting at window transitions as it becomes more sensitive to each update as the lead-time lengthens.

Because joint optimization does not guarantee a unique partition of forecast error between the IC and LoRA corrections, we vary their relative importance to test how the solution depends on that division. The penalty weights are set by a single parameter $\rho$ that is fixed for each joint optimization,

\begin{equation}
\lambda_{\mathrm{IC}}
=
0.3\,e^{-\rho/2},
\qquad
\lambda_W
=
0.3\,e^{\rho/2}.
\label{eq:rho_dial}
\end{equation}

Negative $\rho$ penalizes the initial-condition correction more strongly and shifts the solution toward the model weights, while positive $\rho$ penalizes the model weights more strongly and shifts the solution toward the initial condition. Guided by empirical testing, we evaluate $\rho=\{-20, -16, -12, -8, -4, 0, 4, 8, 12, 16, 20\}$ for each of 36 initialization dates. Finally, the coefficient of 0.3 is chosen based on hyperparameter testing.

\subsection{Gated Optimization}
\label{sec:gated}

The penalty sweep of Section~\ref{sec:joint} sets the relative weighting between the ICs and LoRA, yielding a family of optimal solutions to the shadowing problem. In addition to evaluating GraphCast forecasts to identify the solution within this family that best shadows ERA5, we introduce a second model, Pangu-Weather \citep{bi2023}, to assess how the optimized initial conditions affect its forecasts relative to those initialized from ERA5. We then take this approach one step further, moving from passively validating GraphCast optimization results to playing an active role in the optimization through gated validation. 

Specifically, at each iteration, a proposed IC update is accepted only if it also improves the forecast of Pangu-Weather; otherwise, only the LoRA weights are updated. Pangu-Weather receives only $\delta\mathbf{x}$ and never sees $\delta\mathbf{W}$, so the test is one-sided by construction. Thus, the initial condition accumulates the increments endorsed by Pangu, while the LoRA corrections capture improvements specific to GraphCast. An IC increment that improves both models is more plausibly a correction to the atmospheric state, whereas one that improves only GraphCast is more plausibly specific to GraphCast. If the loss does not decrease, the entire update is rejected.

The optimization minimizes Eq.~\ref{eq:joint_objective} without penalty ($\lambda_{\mathrm{IC}}=\lambda_W=0$), so that the division between the two controls is decided by the referee rather than by a penalty. The LoRA configuration ($r=2$ on all 64 processor matrices) and initialization follow the joint experiment. Each iteration commits at most one update, to either the initial condition or the LoRA. To score a proposed increment, the optimized initial condition is mapped onto Pangu's inputs. At the initialization time, the five upper-air variables common to both models (geopotential, specific humidity, temperature, and the two horizontal wind components) on their 13 shared pressure levels, and the four common surface variables (mean sea-level pressure, 10-m winds, and 2-m temperature), are interpolated from the 1.0$^\circ$ to the 0.25$^\circ$ grid by zero-padded spherical-harmonic transform. The 24-h Pangu-Weather model is then rolled out for 10 days, and the 500-hPa geopotential height ($Z_{500}$) is verified against ERA5. The cosine-latitude-weighted $Z_{500}$ root-mean-square error (RMSE), summed over ten daily leads, is verified against ERA5 and serves as the referee's criterion for accepting updates by comparing each proposed update against the previous best (starting with the control forecast from ERA5). For initialization $i$, this cumulative error is defined as $E_i^{Z_{500}}=\sum_{d=1}^{10}\mathrm{RMSE}_{i,d}^{Z_{500}}$, where $d$ is the forecast lead. A more comprehensive referee loss could be used, but we choose $Z_{500}$ for ease of comparison.

Each initialization date is optimized for 200 iterations over a 5-day window, followed by 200 iterations over a 10-day window. The second window begins from the best-performing state of the first, with freshly initialized optimizers. In order to capture the potentially delayed effects of GraphCast model error embedded in the gated IC, Pangu-Weather evaluates proposed IC updates over the full 10-day forecast, even during the initial 5-day optimization window. The IC and LoRA learning rates both begin at  $5\times10^{-4}$ for the 5-day window. To account for greater sensitivity at longer lead times, the initial learning rates for the 10-day window are halved.

\subsection{Multi-LoRA: Time-Dependent Model Corrections}
\label{sec4:multilora}
Thus far, all LoRA applications define a single weight correction, meaning one $\delta\mathbf{W}$ must serve the entire forecast for one initialization time. The multi-LoRA experiment relaxes this constraint by assigning a separate LoRA correction to each forecast day (4 GraphCast time steps), testing how important time-dependent model error is to the shadowing problem. 

Here we optimize Eq.~\ref{eq:lora_effective_weight} using the procedure described previously, with a separate LoRA adapter applied over each successive four-step period. Specifically, the multi-LoRA adapters are optimized sequentially, one forecast day at a time, with each adapter frozen before proceeding to the next day. Each day uses 30 Adam iterations at a learning rate of $10^{-3}$, minimizing the same weighted forecast MSE while keeping the ERA5 initial condition fixed. Each day's rank-2 LoRA correction follows Eq.~\ref{eq:lora_update} for all 64 processor matrices, with separate factors $\mathbf{B}_{k,j}$ and $\mathbf{A}_{k,j}$ initialized as in Section~\ref{lora_only}. Multi-LoRA uses 1.8 million trainable parameters, compared with 10.8 million for IC-only optimization, but allows corrections throughout the forecast rather than only at the initial time.

\section{Results}\label{sec3:results}
\subsection{Model-Only Optimization via LoRA}
\label{sec:lora_ladder}

\begin{figure}[h]
    \centering
    \includegraphics[width=0.95\columnwidth]{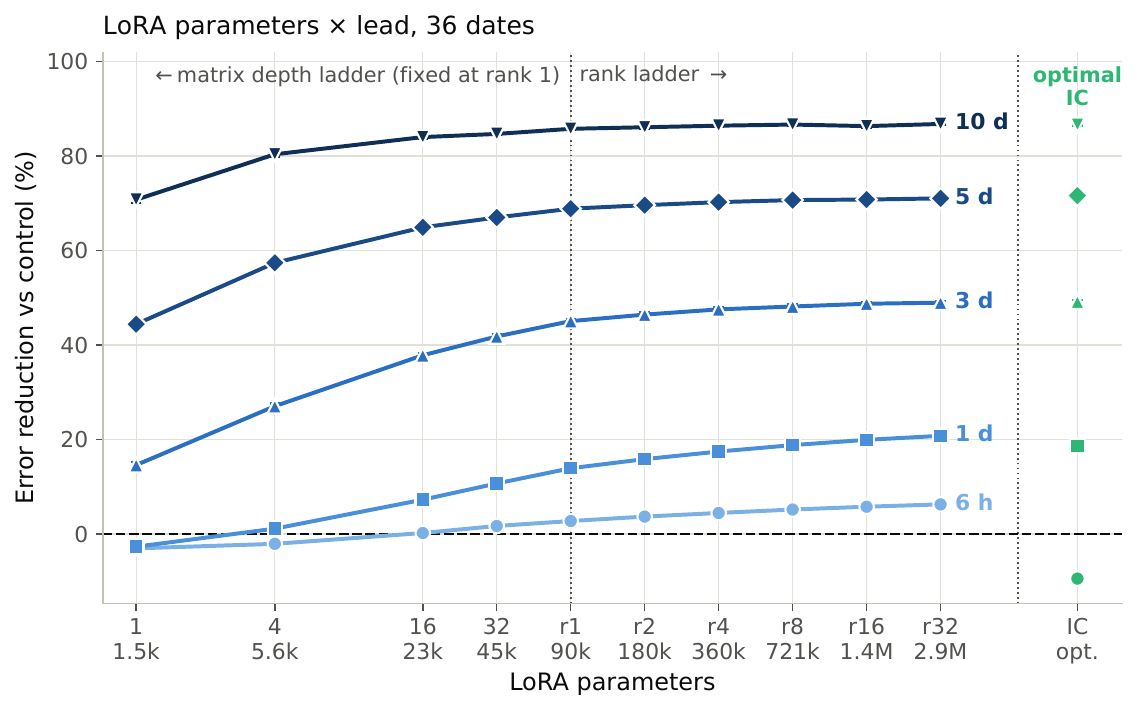}
    \caption{Error reduction (y-axis) relative to the control forecast initialized with ERA5 as a function of the number of trainable LoRA parameters (x-axis) averaged over 36 initialization times. Each curve represents a single lead time. Trainable matrices progressively increase from 1 to all 64 available in the GraphCast processor at fixed rank $r=1$, followed by increasing rank $r$ for all 64 matrices. Error bars show one standard error and are within symbol size. Green symbols show the average improvement for IC optimization using GraphCast for the same cases.}
    \label{fig:lora_ladder}
\end{figure}

In \citet{heatwave_2024} and \citet{vonich2026}, initial conditions were optimized to shadow ERA5 for a static version of GraphCast. Here we solve the complementary problem of optimizing ERA5 10-day shadows for a LoRA-augmented version of GraphCast and a fixed initial condition. We test this across 36 initializations taken from the 1st, 10th, and 20th of each month in 2017, all at 12 UTC. As in IC optimization experiments, the forecast window is progressively extended from 2 to 10 days, carrying the optimized LoRA parameters forward at each extension (see App.~\ref{appendix:baseline_methods} for details).

Figure~\ref{fig:lora_ladder} displays the reduction in forecast error relative to the control at individual lead times (blue curves) as a function of LoRA size. LoRA rank ranges from a maximum of $r=32$ to a minimum of $r=1$ ($2.88\times10^{6}$ to $90{,}112$ parameters). Smaller sizes apply to $r=1$, with the number of adapted processor matrices progressively reduced from 64 to one ($90{,}112$ to $1{,}536$ parameters). For comparison, the full parameter range from 1,536--$2.88\times10^{6}$ equates to about 0.004\%--8.0\% of GraphCast's total trainable parameters, respectively. 

The smallest LoRA networks furnish little benefit at short lead times and can even degrade the early forecast (Fig.~\ref{fig:lora_ladder}), but become increasingly effective as lead time increases. The one- and four-matrix LoRAs, the worst-performing configurations overall, are slightly worse than the control at 6\,h and the four-matrix is neutral at 1\,d. Both, however, still yield improvements at later lead times. Accuracy gains are steepest for all lead times during the increase from 1.5k to the full 90k parameters of the rank-1 LoRA. The largest gain across the full rank sweep occurs at 3\,d lead, with a total improvement of approximately 30\%. The benefit for $r>1$ diminishes after 3\,d, and effectively plateaus at 10\,d. At 10\,d, for $r=1$, increasing the LoRA size from 1--64 matrices yields an increase in error reduction from 71\% to 86\%, whereas increasing rank from $r=1$ to $r=32$ yields a further improvement of only 1\%. By 10 days, the $r=1$ improvements are comparable to IC-only optimization (Fig.~\ref{fig:lora_ladder}, green symbols). 

Doubling the training epochs preserves the overall ordering and rank dependence of the results, except at 6\,h as noted above, while yielding an additional 2--3\% improvement at 10\,d. This indicates that although absolute convergence has not been achieved, there are minimal and diminishing returns from further optimization. Fig.~\ref{fig:appendix_depth} presents the same results with forecast lead shown on the horizontal axis rather than as separate curves, again showing diminishing returns with increasing rank.

\subsection{Joint IC--LoRA Optimization}\label{sec:results_joint}
\begin{figure}[htbp]
    \centering
    \includegraphics[width=0.95\columnwidth]{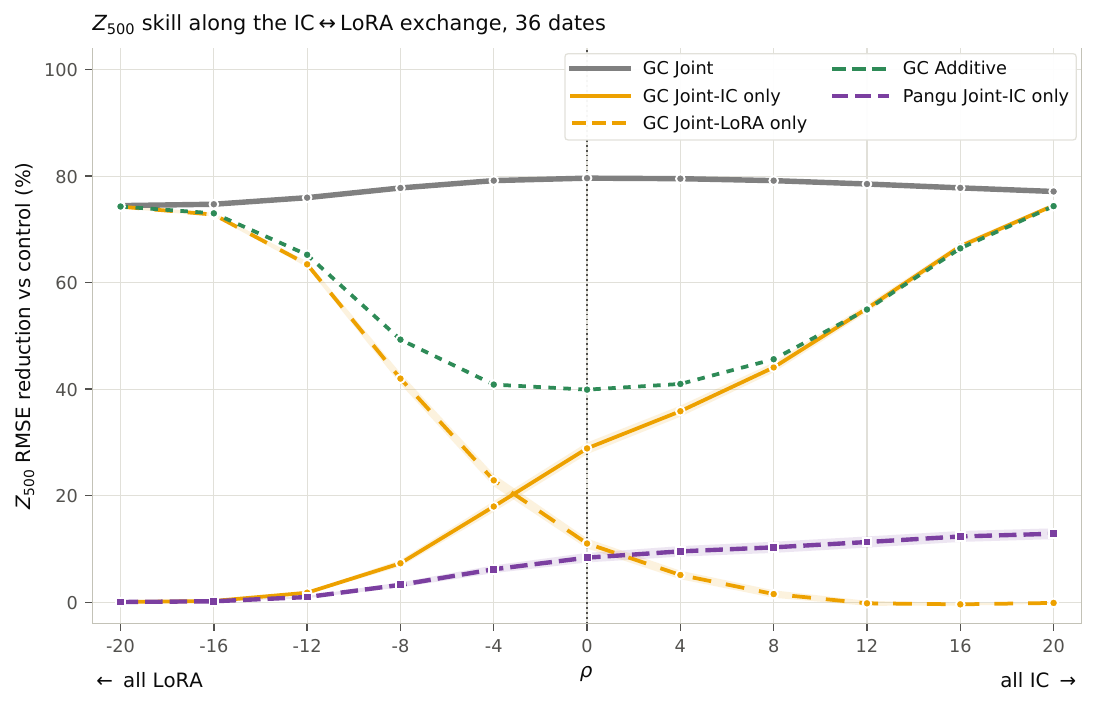}
    \caption{Joint IC--LoRA optimization results as a function of the IC--LoRA weighting term, $\rho$, averaged over 36 initialization dates. The gray curve shows the total 10-day GraphCast Z500 RMSE reduction when both controls are applied. Orange curves show forecast outcome from only using the IC or LoRA component from the joint optimization. The green dashed curve is the sum of the IC-only and LoRA-only reductions. The purple curve shows the Pangu-Weather Z500 RMSE reduction obtained from the optimized IC alone. Shading indicates $\pm1$ standard error of the 36-date mean.}
    \label{fig:joint_x}
\end{figure}

The preceding section demonstrated that model optimization achieves medium-range forecast improvements comparable to those obtained through IC optimization. This parallels the longstanding ambiguity in weak-constraint 4DVAR, where state corrections can compensate for model error and vice versa \citep[e.g.,][]{zupanski1997}. We now investigate the sensitivity of shadowing forecast improvements to the IC--LoRA weighting coefficient $\rho$ (\ref{eq:rho_dial}). Negative $\rho$ penalizes the IC and favors the LoRA solution, while positive $\rho$ favors the IC. For $\rho=0$, the two normalized penalties receive roughly equal weight. For each $\rho$ value, we score four forecasts: (1) the full GraphCast jointly optimized solution, (2) GraphCast with only the IC correction from that joint solution, (3) GraphCast with only the LoRA correction from the joint solution, and (4) Pangu-Weather with only the IC correction from the joint solution. Independently scoring the IC and LoRA components offers insight into their relative contributions and distinguishes the nonlinear interactions that arise when both are included. Evaluating the same IC in Pangu-Weather provides an independent test of whether the state correction generalizes beyond GraphCast. We hypothesize that Pangu-Weather improvements will peak at an intermediate $\rho$, where the optimization can allocate some model-specific error to LoRA rather than absorbing it into the IC corrections.

Figure~\ref{fig:joint_x} reveals that GraphCast forecast skill for the joint solution is insensitive to the relative weighting between the IC and the LoRA. The mean day-1--10 $Z_{500}$ RMSE reduction increases from 74\% at the LoRA-dominated limit ($\rho=-20$) to a broad maximum of about 80\% near $\rho=0$, and then decreases to 77\% at the IC-dominated limit ($\rho=20$). The IC component of the joint optimal increases from about zero improvement over the control for the LoRA limit to 75\% at the IC limit, while the LoRA component falls from 74\% near the LoRA limit to nearly zero at the IC limit. The respective curves cross near $\rho\approx-3$ and their sum (green dashed curve) reveals that the two components do not add up to the joint total (41\% versus 79\% at $\rho=-4$), indicating a nonlinear response in the forecast with both corrections included. 

Contrary to our hypothesis, Fig.~\ref{fig:joint_x} shows that increasing the weight of the jointly-optimized IC leads to monotonic forecast improvements in Pangu-Weather. Mean day-1--10 $Z_{500}$ RMSE reduction rises from zero at $\rho=-20$ to 13\% at $\rho=20$, with no local maximum. Improvements diminish toward the IC limit: as $\rho$ increases from 0 to 20, GraphCast IC-only skill more than doubles from 29\% to 75\%, while Pangu-Weather skill increases only modestly, from 8\% to 13\%. Thus, additional IC weighting remains beneficial to Pangu, but an increasing fraction of the added GraphCast skill does not transfer.

\begin{figure}[t]
    \centering
    \includegraphics[width=0.95\columnwidth]{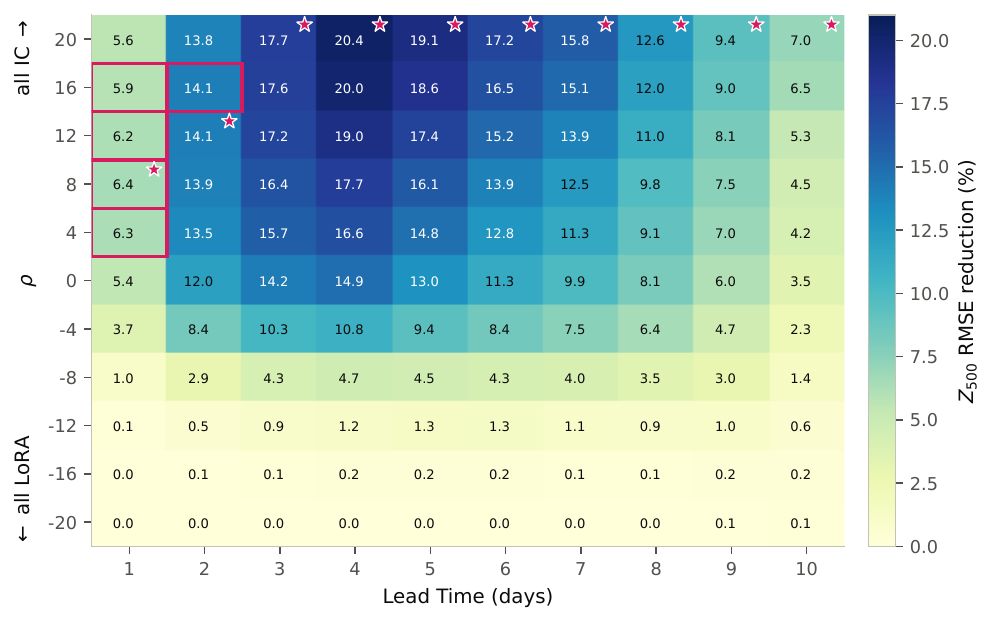}
    \caption{Performance of the GraphCast joint solution's IC component from Fig.~\ref{fig:joint_x} in Pangu-Weather as a function of forecast lead and $\rho$. Each cell shows the mean reduction in $Z_{500}$ RMSE relative to each date's Pangu-Weather control averaged over 36 cases. Red boxes indicate $\rho$-values that outperform $\rho=20$ at that lead under a paired two-sided $t$-test ($p<0.01$). Red stars mark the best-performing $\rho$ at each lead time.}
    \label{fig:heatmap}
\end{figure}

Examining the performance of the IC corrections in Pangu-Weather as a function of the weighting parameter $\rho$ and forecast lead time shows that relative improvement peaks near day~4 in Pangu-Weather for all $\rho$ (Fig.~\ref{fig:heatmap}). This is similar to the behavior found in \citet{vonich2026} when testing GraphCast-optimized ICs in Pangu. The one exception to the monotonic dependence of forecast skill on $\rho$ occurs at day~1, and more weakly at day~2, where Pangu-Weather improvements peak at an intermediate $\rho$ rather than at the IC-dominated limit. In other words, Pangu-Weather benefits from allocating more of the correction to LoRA at short leads, but this advantage is gone by day~3.

\subsection{Cross-Model Validation of Initial-Condition Updates}
\label{sec:gating}
\citet{laloyaux2020exploring} show that weak-constraint 4DVAR can partially separate analysis and model error in one forecast system by assigning \textit{a priori} spatial correlation scales to each. Their results highlight that separating the two sources requires additional information beyond forecast verification alone. The joint optimization experiment is consistent with that finding, since similar forecast improvements exist across the full spectrum of IC--LoRA weighting. Since solutions from the joint optimization are underdetermined, we introduce Pangu-Weather as an external referee on the IC correction, yielding a gate the optimization procedure must satisfy. Specifically, we repeat the joint optimization algorithm with the following adjustment. 

At each iteration, GraphCast computes gradients of the forecast loss with respect to both the IC and LoRA controls. Each proposed IC update is evaluated against the summed day-1--10 $Z_{500}$ error in GraphCast and Pangu-Weather, yielding four possible outcomes:
(1) if both model forecasts improve, the IC update is accepted;
(2) if only GraphCast's forecast improves, the IC update is rejected and the LoRA update is accepted;
(3) if only Pangu-Weather's forecast improves, neither the IC nor the LoRA is accepted; and
(4) if the loss increases, the search continues without an update.

As before, because LoRA modifies only GraphCast, Pangu-Weather receives only the IC correction. The cross-model gate thus restricts IC updates to those benefiting both models while allowing GraphCast-specific improvements to be captured by LoRA. We note that model errors shared by GraphCast and Pangu-Weather may still pass through the gate into the initial condition. We hypothesize that this approach will produce more comparable IC-only improvements in the two models while preserving additional GraphCast skill through LoRA. We compare gated and ungated optimizations for the same 36 initialization dates used in Fig.~\ref{fig:joint_x}. The principal difference between the ungated and gated experiments is the acceptance criterion: ungated optimization retains GraphCast-improving IC updates regardless of their effect on Pangu, as in \citet{vonich2026}.

During the gate experiment, we begin by optimizing the 5-day trajectory and then expand the window to 10 days. We choose 5 days as the natural breakpoint because of the 4-day Pangu-Weather performance peak in Fig.~\ref{fig:heatmap}. Of the 200 proposed updates for each initialization date during the 5-day window, 17\% improved forecasts for both models and were committed to the IC, 45\% were committed to LoRA, and the remaining 38\% produced no loss reduction and were skipped. In the subsequent 10-day window the corresponding shares were 10\%, 22\%, and 68\%, respectively. As the optimization window increases, fewer proposals are accepted by either channel. Rejections concentrate at later epochs for each window, with most dates eventually entering a terminal stall in which little to no additional benefit is seen for Pangu-Weather (see Fig.~\ref{fig:indiv_cases}). We expect this is partially due to our relatively large learning rate, which produces noisier updates but more clearly exposes directions that improve GraphCast while degrading Pangu. In this sense, gating acts as a selective form of early stopping for the IC with respect to Pangu. A refined learning-rate schedule reduced stalling in a subset of experiments and reached comparable skill more gradually, without significant additional improvement (not shown).

\begin{figure}[t]
\centering
\includegraphics[width=.95\textwidth]{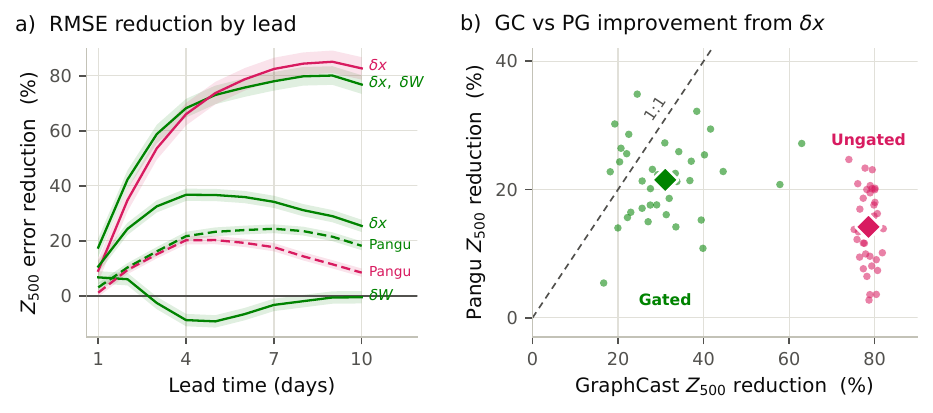}
\caption{Comparison of forecast error improvements for the gated and ungated optimizations for all 36 initialization dates (green denotes gated; red, ungated). (a) Average $Z_{500}$ forecast error reduction by lead time. Solid curves show GraphCast error reduction from the gated solution with the IC ($\delta x$) and LoRA ($\delta W$) corrections applied independently or jointly ($\delta x, \delta W$). Dashed curves show the forecast improvement of gated and ungated $\delta x$ for Pangu-Weather. Shading denotes $\pm1$ standard error. (b) Per-date reduction in Pangu-Weather $Z_{500}$ error from the gated $\delta x$ alone, plotted against the reduction produced by the same $\delta x$ in GraphCast. Markers represent the forecast error reduction of the summed day-1--10 RMSE relative to each model's own corresponding control. Bold diamonds designate the 36-initialization means.}
\label{fig:gate_composition}
\end{figure}

Figure~\ref{fig:gate_composition}a shows the mean $Z_{500}$ RMSE reduction by forecast lead in GraphCast and Pangu-Weather for the gated and ungated experiments (see also Fig.~\ref{fig:app_gated_T850} for $T_{850}$; and Fig.~\ref{fig:app_gated_U250} for $U_{250}$). Forecast performance is also assessed for the individual components of the joint optimization. Without gating, GraphCast error reduction from $\delta x$ alone is 79\%, compared with 31\% with gating, bringing it substantially closer to Pangu-Weather's 21\% reduction (up from 14\% without gating). The 10\% gating performance gap between GraphCast and Pangu-Weather may represent model-specific error that evades the gate, though Pangu's better baseline may also account for part of the difference---its matched controls have, on average, 14\,m less RMSE than GraphCast to begin with. Interestingly, the GraphCast forecast improvements excluded from the state correction are not lost. When the gated IC and LoRA correction are applied together, they reduce GraphCast error by 75\%, on par with the 79\% obtained from the ungated IC and in line with all joint $\rho$ combinations in Fig.~\ref{fig:joint_x}. The LoRA correction yields very little improvement on its own. Applied to the unmodified ERA5 initial condition, it enhances GraphCast forecasts by 7\% on the first day and degrades them by 9\% on day~5. This indicates that the LoRA is strongly co-adapted to its associated $\delta x$, and there is significant nonlinearity in the IC-LoRA combination. This may reflect state-specific model error, or it could be that the LoRA component is a potent reservoir for overfitting.  

Figure~\ref{fig:gate_composition}b illustrates how error reduction varies between the two models for each initialization.  As we have seen, ungated ICs produce large improvements in GraphCast but substantially smaller gains in Pangu, whereas gated ICs produce broadly comparable improvements across the two models. Specifically, the mean forecast performance gap shrinks from 64\% without gating to 10\% with gating. Notably, several gated ICs improve Pangu-Weather more than they do GraphCast, while none of the ungated ICs have this property.

\subsection{Time-Dependent Model Correction}\label{results:multilora}

\begin{figure}[t]
    \centering
    \includegraphics[width=0.9\columnwidth]{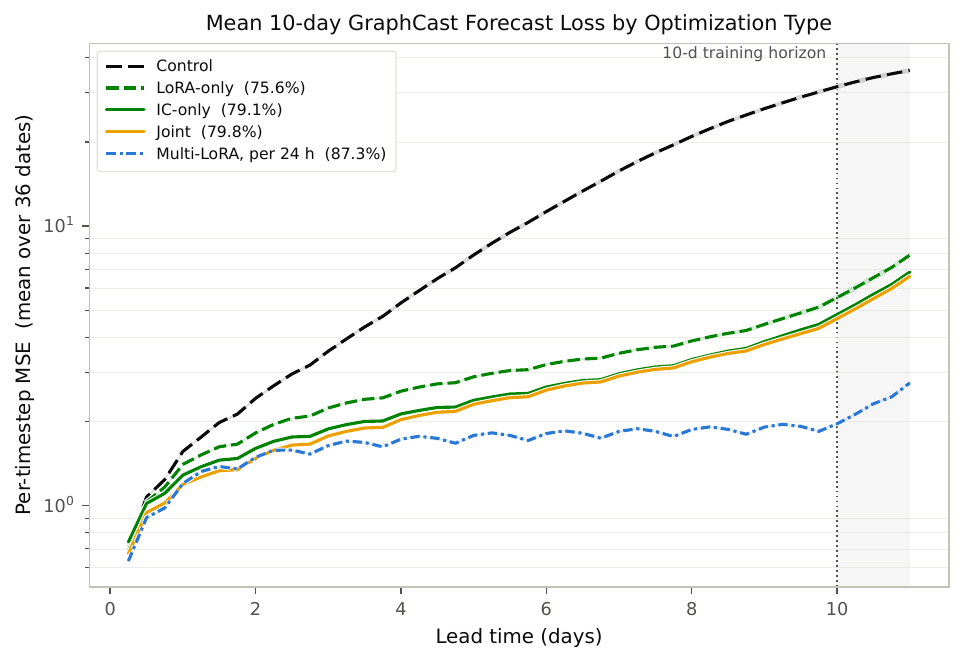}
    \caption{Mean GraphCast forecast loss by lead time evaluated every 6~h on a logarithmic vertical axis. The control uses unmodified ERA5 initial conditions and GraphCast weights. IC-only optimizes only the initial conditions; LoRA-only optimizes a single set of rank-2 matrices for each initialization; and Joint optimizes ICs and a single-set of rank-2 matrices simultaneously. Multi-LoRA fits a separate set of rank-2 matrices for each 24\,h period for all initialization times. Legend percentages show reductions in loss relative to controls over the whole 10-day window. The shaded region beyond day~10 lies outside the optimization window, where multi-LoRA applies no adapters.}
    \label{fig:multilora}
\end{figure}

In a final experiment, we test whether the similar performance of IC-only, LoRA-only, and joint optimization reflects a practical limit imposed by corrections that, once optimized, cannot change as the forecast evolves. Figure~\ref{fig:multilora} compares these approaches across the same 36 initializations, showing a mean reduction of GraphCast forecast loss over ten days as 79\%, 76\%, and 80\%, respectively. These approaches impose relatively restrictive, physically motivated controls: the initial state is adjusted only at initialization and any model-weight correction remains fixed throughout the forecast. Multi-LoRA relaxes this constraint by fitting a separate rank-2 model correction for each 24\,h forecast period, achieving a larger error reduction of 87\%. This time-dependent correction is conceptually closer to weak-constraint 4DVAR, which permits model-error adjustments throughout the trajectory, although multi-LoRA modifies the forecast operator rather than applying additive forcing. Its substantially greater flexibility in time makes trajectory-specific overfitting more plausible. Some degree of overfitting may also occur under the more constrained approaches, and we cannot determine in this study where physically meaningful correction ends and trajectory-specific fitting begins for any of them. The purpose of this figure is therefore not to identify the most accurate representation of atmospheric error, but to demonstrate that each class of correction has an apparent performance limit.

\section{Discussion and Conclusions}\label{sec5:conclusion}
We introduce adaptive model correction by training LoRA adjustments for the GraphCast model to better shadow ERA5, with the aim of reducing the inclusion of model error in optimized initial conditions. GraphCast optimizations modifying either the initial state or the model in isolation recover similar amounts of forecast skill improvement over an unmodified control. Optimizing a loss function that contains both the initial conditions and LoRA parameters together achieves comparable forecast improvements despite introducing both correction pathways at once. We find this to be the case for the entire range of IC--LoRA weighting. These experiments could not, therefore, clearly identify error as belonging to the state versus the model.

Cross-model evaluation provided a partial constraint on the degeneracy that affects the joint IC--LoRA experiments. Online gating with Pangu-Weather restricted the GraphCast improvement from IC correction to a level comparable to that in Pangu-Weather, while also increasing the mean Pangu-Weather forecast improvement relative to ungated optimization (Fig.~\ref{fig:gate_composition}). The GraphCast forecast improvements excluded from the IC are redirected to a LoRA update, preserving the portion of the update that addresses GraphCast-specific model error. Without early stopping or gating, continued GraphCast-focused IC optimization can eventually degrade Pangu-Weather forecasts, even as GraphCast improves.

However, the IC and LoRA corrections do not act independently. Their individual forecast improvements are not additive, and this interaction is especially apparent in the gated experiment, where LoRA contributes substantially to the joint forecast improvement but provides little benefit when applied independently of its associated optimal IC. This suggests that the two corrections become strongly coupled through nonlinear co-adaptation during optimization, in contrast to the additive model-error formulation of traditional weak-constraint 4DVAR. Consequently, the LoRA correction cannot necessarily be interpreted as an independent estimate of GraphCast model error, but may instead represent a state-specific correction or a source of overfitting.

Although cross-model gating constrains IC optimization, agreement between models does not necessarily imply a more accurate atmospheric state. Shared model errors and biases in ERA5 may still influence the optimized ICs. Improved cross-model agreement provides evidence of greater generalization but cannot prove superior analysis accuracy, which would require further validation against independent observations and additional forecast models.

The time-dependent LoRA experiment extends shadowing beyond the apparent limits of fixed IC and model corrections, but with greater potential for trajectory-specific overfitting. This raises uncertainty about our earlier interpretation of 30-day shadowing as evidence for a longer intrinsic predictability limit \citet{vonich2026}, even though it was obtained without time-dependent model corrections. Where does meaningful shadowing end and overfitting begin? Answering this question is an important direction for future research.

Gated optimization establishes a practical starting point for more advanced state--model error separation. One extension could use two fully differentiable models to optimize the same IC together, allowing cross-model agreement to constrain state corrections directly rather than through an external accept--reject step. Although Pangu-Weather's pretrained weights are publicly available, its official implementation supports inference but not automatic differentiation, necessitating the gating approach used here. Future work could also extend gating to the LoRA correction itself, potentially helping identify reanalysis biases. Together, these approaches offer a path toward more reliable separation of state and model error, supporting more physically interpretable forecast-error corrections.

%%%%%%%%%%%%%%%%%%%%%%%%%%%%%%%%%%%%%%%%%%%%%%%%%%%%%%%%%%%%%%%%%%%%%
% ACKNOWLEDGMENTS
%%%%%%%%%%%%%%%%%%%%%%%%%%%%%%%%%%%%%%%%%%%%%%%%%%%%%%%%%%%%%%%%%%%%%
\section*{Acknowledgments}
We acknowledge high-performance computing support from the Casper cluster (\url{https://doi.org/10.5065/qx9a-pg09}) provided by NCAR's Computational and Information Systems Laboratory, sponsored by the National Science Foundation. The Copernicus Climate Data Store provided access to ERA5. We thank Ryan Keisler and Hans Mohrmann (Brightband) for conversations related to this work. This research was supported by grants 2023-4715 from the Heising-Simons Foundation and 2501400 from the National Science Foundation. Anthropic and OpenAI products were used for assistance with plotting code, experimental design, and manuscript editing.

%%%%%%%%%%%%%%%%%%%%%%%%%%%%%%%%%%%%%%%%%%%%%%%%%%%%%%%%%%%%%%%%%%%%%
% APPENDIX
%%%%%%%%%%%%%%%%%%%%%%%%%%%%%%%%%%%%%%%%%%%%%%%%%%%%%%%%%%%%%%%%%%%%%
\clearpage
\appendix
\setcounter{figure}{0}
\renewcommand{\thefigure}{A\arabic{figure}}

\section{GraphCast Model and Initial-Condition Optimization}
\label{appendix:baseline_methods}

The GraphCast configuration, initial-condition optimization procedure, and forecast loss function used throughout this study are identical to those described in \citet{vonich2026} except for the specific updates stated in Sec.~\ref{sec2:data_methods}. The reused methods are repeated here for completeness and reproducibility.

\subsection*{Model}
We use the ``small'' version of the GraphCast model \citep{lam2023learning}, selected for its modest memory footprint. This enables gradient computations over extended windows, up to 32 days, in a practical timeframe. Since this configuration operates on a 1.0° grid, it does not resolve mesoscale processes. As a result, our findings apply only to predictability at synoptic and planetary scales.

GraphCast forecasts six atmospheric state variables: geopotential, temperature, specific humidity, vertical velocity, and zonal and meridional wind components, resolved across 13 pressure levels. It also predicts four surface variables—mean sea-level pressure, 2-meter air temperature, and 10-meter zonal and meridional wind components—alongside 6-hour accumulated precipitation, all on a 1.0° × 1.0° grid. With 36.7 million parameters, GraphCast was trained on ERA5 reanalysis data from 1979 to 2015 and validated on 2016/2017 \citep{hersbach2017era5}. None of the forecasts optimized in this study are part of the training data.

To generate predictions, inference utilizes two atmospheric input states, separated by a 6-hour interval, producing a single output state 6 hours in the future. For extended forecasts, the output is autoregressively fed back into the model alongside the prior 6-hour state, enabling indefinite prediction. In this study, optimal perturbations are computed exclusively for the state variables, while static fields, including the land-sea mask and surface geopotential, remain unaltered.

\subsection*{Optimization}
Our approach leverages the fully differentiable nature of GraphCast to optimize initial conditions in a nonlinear framework, overcoming the limitations of traditional adjoint models \citep[e.g.,][]{langland1995,errico1997adjoint} related to computational expense. Machine-learning models integrate linear and nonlinear operations across layers, enabling seamless derivative computation via the chain rule. In this study, all automatic differentiation is performed using GraphCast implemented in the JAX framework \citep{lam2023learning}. JAX provides robust support for automatic differentiation, complemented by GPU acceleration and dynamic code optimization \citep{jax2018github}.

This differentiable framework enables iterative refinement of atmospheric initial conditions. Given the input state ${\bf x}_i$ for iteration $i$ (where $i=0$ is the unaltered ERA5 initial condition), we compute an increment based on the gradient of the forecast loss, $\mathcal{L}({\bf N}({\bf x}_i))$, with respect to the inputs. Here, ${\bf N}$ denotes GraphCast autoregressive inference beginning from the chosen forecast optimization time. For illustration, the corresponding gradient-descent update is

\begin{equation}
{\bf x}_{i+1} = {\bf x}_i - \eta\frac{\partial \mathcal{L}}{\partial {\bf x}_i}
\label{eqn:increment}
\end{equation}

\noindent The derivative in (\ref{eqn:increment}) entails tracing the loss gradient back through the GraphCast neural network for every 6-hour time step. This gradient characterizes how changes to the inputs influence the loss function. Adjoint models can also be employed to perform gradient descent, but, like deep learning models, they may struggle to navigate complex gradient landscapes with numerous saddle points and valleys \citep{Pires1996}. This phenomenon appears in forecast optimizations beyond 5 days due to increasing gradient complexity with longer lead times \citep{heatwave_2024}. To overcome this problem, the optimization window is gradually expanded rather than fitting the entire trajectory at once. This simplifies the initial optimization and allows the solution to be refined over progressively longer forecast windows. The window lengths and iteration counts vary by experiment, as specified in the main Methods section. \citet{Swanson-1998} implement a similar strategy for a four-dimensional variational data assimilation solver, also noting the performance improvement offered by progressively assimilating the total available data. They refer to this method as quasi-static, reflecting the stepwise adjustment of the assimilation window.

Details of the algorithm used to produce one set of optimized initial conditions (an ``optimal'') are as follows:
\begin{enumerate}
\item Given an initial state, produce a forecast over the prescribed optimization window. The first iteration begins from the ERA5 analysis, while subsequent iterations use the updated initial conditions.
\item Calculate the forecast loss by verifying against ERA5 at every forecast step within the window.
\item Calculate the gradient of the loss with respect to the two input times ($t$ and $t-6\,\mathrm{h}$) using JAX.
\item Update the inputs using the Adam optimizer \citep{kingma2017}, following the gradient direction illustrated in Eq.~\ref{eqn:increment}.
\item Repeat steps 1--4 for the prescribed number of iterations, then extend to the next optimization window using the selected solution from the preceding window. Continue until the final window is reached.
\end{enumerate}

The experiments in this study use optimization windows extending to a maximum of 10 days. Longer windows increase the memory required to compute gradients through the autoregressive forecast, motivating the use of the 1.0$^\circ$ GraphCast configuration. Optimization uses Adam with default momentum parameters ($\beta_1=0.9$, $\beta_2=0.999$), while learning rates vary by experiment as specified in the main Methods section. Each initialization is optimized separately, without batching. For variables constrained to non-negative values, such as specific humidity and precipitation, optimization occasionally produces small negative values; these are clipped to zero with negligible impact on the results. All forecasts using optimized initial conditions are evaluated after this clipping.

\subsection*{Loss Function}
As in \citet{heatwave_2024}, we adopt the scalar loss function used to train GraphCast, a weighted mean squared error (MSE) that quantifies the difference between predicted and target outputs, averaged over time, variables, and spatial locations. For a predicted state $\hat{x}$ and verification state $x$, the loss is expressed as:

\begin{equation}
\label{eq2:loss}
\resizebox{0.7\columnwidth}{!}{%
$\displaystyle
\mathcal{L}
=
\underbrace{
\frac{1}{T_{\mathrm{time}}}
\sum_{\tau=1}^{T_{\mathrm{time}}}
}_{\text{lead time}}
\;
\underbrace{
\frac{1}{|G_{1.0^\circ}|}
\sum_{i\in G_{1.0^\circ}}
}_{\text{spatial location}}
\;
\underbrace{
\sum_{j\in J}
}_{\text{variable-level}}
s_j w_j a_i
\left(
\hat{x}_{i,j}^{t_0+\tau}
-
x_{i,j}^{t_0+\tau}
\right)^2
$%
}
\end{equation}

\noindent In this equation, $w$ represents the weight by pressure level, $a$ is the grid-cell area, and $s$ is a standardization parameter computed from time differences in the GraphCast training data. In \citet{vonich2026}, we demonstrate that the MSE optimization does not merely smooth its way to late-trajectory skill, but genuinely reduces information error. For more details on these variables and the loss function, refer to Section 4.2 of the original GraphCast paper \citep{lam2023learning}.

\section{Additional Figures}

\begin{figure}[t]
    \centering
    \includegraphics[width=\columnwidth]{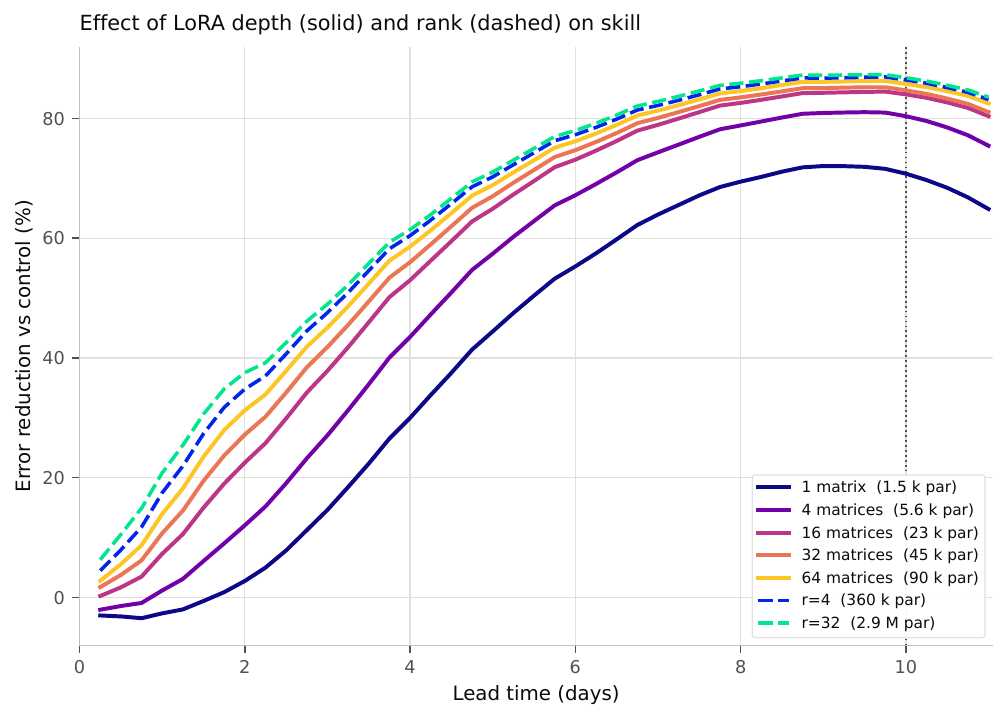}
    \caption{LoRA depth versus rank presented in an alternative form to Fig.~\ref{fig:lora_ladder}. GraphCast objective loss reduction is relative to the control forecast as a function of lead time for LoRA-only optimization out to ten days (averaged over 36 dates). Solid curves represent the matrix depth varied at fixed rank~1, adapting the first 1, 4, 16, 32, or all 64 processor weight matrices. Dashed curves represent all 64 matrices at rank~4 and rank~32. The legend gives trainable parameter counts.}
    \label{fig:appendix_depth}
\end{figure}

\begin{figure}[p]
    \centering
    \includegraphics[
        width=\columnwidth,
        height=0.815\textheight,
        keepaspectratio
    ]{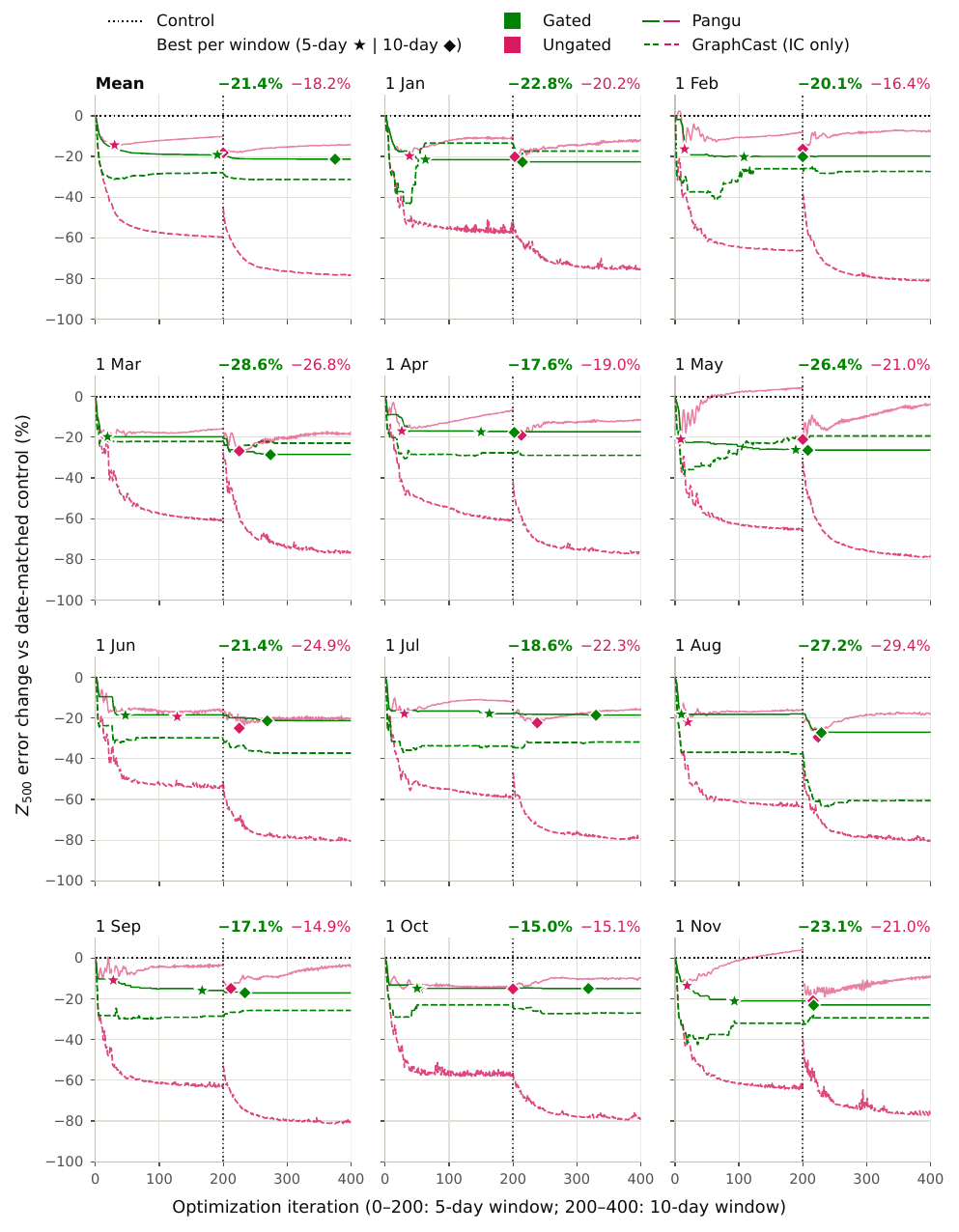}
\caption{Gated and ungated optimizations evaluated in GraphCast and Pangu-Weather. The first panel (upper-left) shows the mean over the 36 dates; remaining panels show first-of-month 12~UTC optimization descents from January--November 2017. The y-axis metric is the summed day-1--10 $Z_{500}$ RMSE verified against ERA5, expressed as percent change from each model's date-matched control. The gated arm optimizes the initial conditions jointly with a LoRA weight correction; its GraphCast curve uses the initial conditions alone (LoRA removed), the same state Pangu receives. Stars and diamonds mark the lowest Pangu errors in the 5- and 10-day windows, respectively. Percentages above each panel show the best Pangu improvement relative to control for gated (green) and ungated (red) optimization. Ungated outperforms gated on half of dates when stopped early. Gating does not guarantee a better Pangu performance, but it does produce more comparable GraphCast and Pangu-Weather improvements on all 36 cases.}
\label{fig:indiv_cases}
\end{figure}

\begin{figure}[t]
\centering
\includegraphics[width=\textwidth]{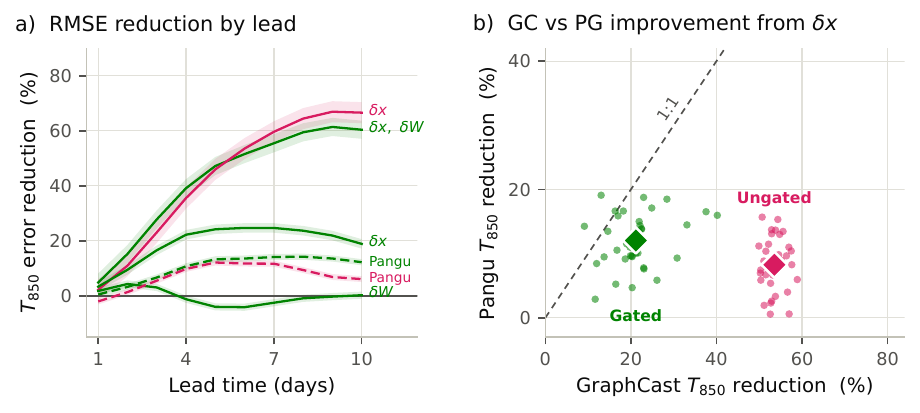}
\caption{As in Fig.~\ref{fig:gate_composition}, but measuring 850~hPa temperature.}

\label{fig:app_gated_T850}
\end{figure}

\begin{figure}[t]
\centering
\includegraphics[width=\textwidth]{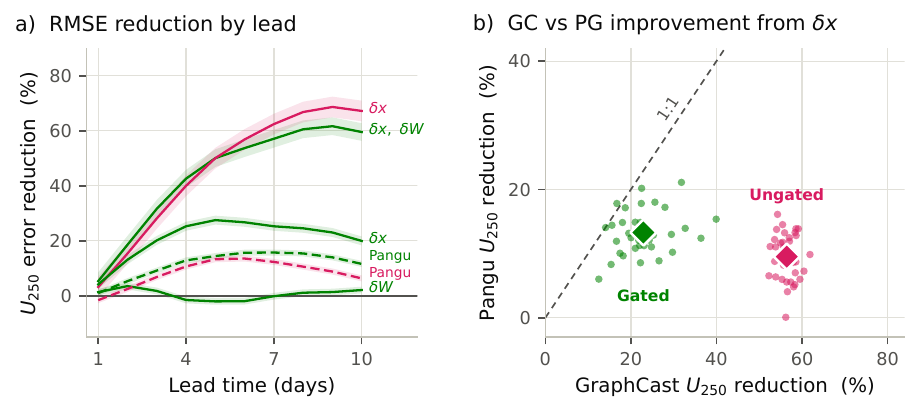}
\caption{As in Fig.~\ref{fig:gate_composition}, but measuring 250~hPa wind.}

\label{fig:app_gated_U250}
\end{figure}

\clearpage

\bibliographystyle{plainnat}  
\bibliography{references}

\end{document}